\documentclass[aps,preprint,showpacs,preprintnumbers,amsmath,amssymb]{revtex4}

\usepackage{dcolumn}
\usepackage{mathrsfs}% Align table columns on decimal point
\usepackage{bm}% bold math
\usepackage{amsmath,amssymb,epsfig,float}

\begin{document}

\title{Effects of prepared states and Unruh temperature  on Measurement-Induced-Nonlocality}

\author{Zehua Tian and Jiliang Jing\footnote{Corresponding author, Email: jljing@hunnu.edu.cn}}
\affiliation{Department of Physics, and Key Laboratory of Low
Dimensional Quantum Structures and Quantum
Control of Ministry of Education,\\
Hunan Normal University, Changsha, Hunan 410081, China}

\vspace*{0.2cm}
\begin{abstract}
\vspace*{0.2cm} How the prepared states and Unruh effect  affect
Measurement-Induced-Nonlocality (MIN) is studied. We show that, as
the Unruh temperature increases, the MIN between modes $\mathrm{A}$
and $\mathrm{I}$ decreases but the MIN  between modes $\mathrm{A}$
and $\mathrm{II}$ increases. We prove that the parameters $c_i$
which decide initial prepared states affect not only the values of
the MIN, but also the dynamical behavior of it. By comparing the MIN
with the maximal expectation values of CHSH inequality and geometric
discord between modes $\mathrm{A}$ and $\mathrm{I}$, we also find
that the MIN is more general than the quantum nonlocality related to
violation of Bell's inequalities, and its values is always equal or
bigger than that of the geometric discord.
\end{abstract}

\vspace*{1.5cm}

\maketitle

\section{introduction}

The investigation of relativistic quantum information not only
supplies the gap of interdiscipline refer to quantum information and
relativity theory, but also has a positive promotion on the
development of them. As a result of that, this domain has been paid
much attention in the last decade \cite{Alsing1,Alsing2,Fuentes,Yi
Ling,Shahpoor,Pan,Montero,Wang,Datta,Han,David,Mann}. Among them,
most papers have focused on quantum resource, e. g., quantum
entanglement\cite{Alsing1,Alsing2,Fuentes,Yi
Ling,Shahpoor,Pan,Montero,Mann} and discord \cite{Wang,Datta},
because quantum resource plays an very important role in the quantum
information tasks such as teleportation \cite{Bennett} and
computation \cite{Horodecki,Bouwmeester}, and studying it in a
relativistic setting is very closely related to the implementation
of quantum tasks with observers in arbitrary relative motion. In
addition, extending this work to the black hole background is very
helpful for us to understand the entropy and paradox
\cite{Lee,Hawking} of the black hole.

Despite much effort has been paid to extend quantum information
theory to the relativistic setting, another important foundation of
quantum mechanics---nonlocality is barely considered. Recently,
Nicolai Friis $\emph{et al}$ firstly studied the nonlocality in the
noninertial frame, and they pointed out that residual entanglement
of accelerated fermions is not nonlocal \cite{Nicolai}. Following
them Alexander Smith $\emph{et al}$ studied the tripartite
nonlocality in the noninertial frames \cite{Smith}, and DaeKil Park
considered tripartite entanglement-dependence of tripartite
nonlocality \cite{Park}. Generally, most researchers analyzed the
quantum nonlocality by means of Bell's inequalities \cite{J. F.
Clauser} for bipartite system and Svetlichny inequality for
tripartite system \cite{G. Syetlichny}, respectively. Because, these
inequalities are satisfied by any local hidden variable theory, but
they may be violated by quantum mechanics. However, Shunlong Luo and
Shuangshuang Fu have introduced a new way to quantify nonlocality by
measurement, which is called the MIN \cite{Luo}, and following their
paper, a number of papers emerged to perfect its definition
\cite{Xi,Sayyed} and discussed its properties \cite{Sen,Guo}. In
addition, some authors have analyzed its dynamical behavior and
compared it with other quantum correlation measurements such as the
geometric discord \cite{Hu,Zhang}. However, all of these studies
don't involve the effect on the MIN resulting from relativistic
effect. In fact, the study that how the Unruh effect \cite{Unruh}
affects the MIN not only is very significant to theory study, but
also plays a key role in practice, which can help us to implement
the quantum task preferably and more efficiently. Inspired by that,
we analyze how the Unruh effect and prepared states affect the MIN
in this paper and we will show some new properties.

Our paper is constructed as follows. In section II we introduce  the
different vacuums for relativistic observers and the definition of
the MIN. In section III how the Unruh effect and prepared states
affect the MIN are studied. And  in the last section we summarize
and discuss our conclusions.

\section{Vacuums, excited sates and definition of MIN }

It is well known that an uniformly accelerated observer  will detect
a thermal particle distribution in the Minkowski vacuum. The
Minkowski vacuum can be factorized as a product of the vacua of all
different Unruh modes
\begin{eqnarray}\label{Minkowsik vacuum}
|0\rangle_\mathrm{M}=\otimes_w|0_w\rangle_\mathrm{U}.
\end{eqnarray}

For the Dirac field, the Unruh monochromatic mode
$|0_w\rangle_\mathrm{U}$, from the noninertial observers'
perspective, can be expressed as \cite{Alsing2,Montero,Wang}
\begin{eqnarray}\label{vacuum}
|0_w\rangle_\mathrm{U}=(e^{-w/T}+1)^{-\frac{1}{2}}|0_w\rangle_\mathrm{I}|0_w\rangle_\mathrm{II}
+(e^{w/T}+1)^{-\frac{1}{2}}|1_w\rangle_\mathrm{I}|1_w\rangle_\mathrm{II},
\end{eqnarray}
where $|m\rangle_\mathrm{I}$ $(|n\rangle_\mathrm{II})$  denote
Rindler mode in region $\mathrm{I}$ (region $\mathrm{II}$), and
$\mathrm{T}=a/2\pi$ is the Unruh temperature in which $a$ denotes
the proper acceleration of the noninertial observer. Likewise, the
particle state of Unruh mode $w$ in the Rindler basis is found to be
\begin{eqnarray}\label{particle state}
|1_w\rangle_\mathrm{U}=|1_w\rangle_\mathrm{I}|0_w\rangle_\mathrm{II}.
\end{eqnarray}

Recently, Luo $\emph{et al}$ \cite{Luo} have introduced a way  to
quantify nonlocality from a geometric perspective in terms of
measurements, which is named the MIN. For a bipartite quantum state
$\rho$ shared by subsystem $A$ and $B$ with respective to Hilbert
space $H^A$ and $H^B$, we can find the difference between the
overall pre-measurement and post-measurement states by performing a
local von Neumann measurements on part $A$. To capture the genuine
nonlocal effect of the measurements on the state, the key point is
that the measurements do not disturb the local state
$\rho^A=tr_B\rho$. Based on this idea, the MIN can be defined by
$\emph{et al}$ \cite{Luo}
\begin{eqnarray}\label{intial MIN}
N(\rho)=\max_{\Pi^A}\parallel\rho-\Pi^A(\rho)\parallel^2.
\end{eqnarray}
For a general $2\times2$ dimensional system
\begin{eqnarray}\label{Bloch representation}
\rho=\frac{1}{2}\frac{\mathbf{1}^A}{\sqrt{2}}\otimes\frac{\mathbf{1}^B}{\sqrt{2}}
+\sum^3_{i=1}x_iX_i\otimes\frac{\mathbf{1}^B}{\sqrt{2}}
+\frac{\mathbf{1}^A}{\sqrt{2}}\otimes\sum^3_{j=1}y_jY_j
+\sum^3_{i=1}\sum^3_{j=1}t_{ij}X_i\otimes Y_j,
\end{eqnarray}
its MIN is given by \cite{Luo}
\begin{eqnarray}\label{MIN}
N(\rho)=
\left\{
\begin{array}{lr}
trTT^t-\frac{1}{\|\mathbf{x}\|^2}\mathbf{x}^tTT^t\mathbf{x}  & ~~{\text{if}}~~ \mathbf{x}\neq0,\\
trTT^t-\lambda_3  & ~~{\text{if}}~~ \mathbf{x}=0,
\end{array}
\right.
\end{eqnarray}
where $TT^t(T=(t_{ij}))$ is a $3\times3$ dimensional matrix,
$\lambda_3$ is its minimum eigenvalue, and
$\|\mathbf{x}\|^2=\sum_ix^2_i$ with $\mathbf{x}=(x_1,x_2,x_3)^t$.

\section{MIN of X-type initial states}

We now assume that Alice and Rob share a X-type initial state
\begin{equation}\label{initial states}
\rho_{AB}=\frac{1}{4}\left(I_{AB}+ \sum_{i=1}^{3}c_{i}\sigma_{i}%
^{(A)}\otimes\sigma_{i}^{(B)}\right),
\end{equation}
where $I_{A(B)}$ is the identity operator in subspace $A(B)$,  and
$\sigma_{i}^{(n)}$ is the Pauli operator in direction $i$ acting on
the subspace  $n=A,B$, $c_{i} \in\mathfrak{R}$ such that $0\leq \mid
c_{i}\mid\leq1$ for $i=1,2,3$. Obviously, Eq. (\ref{initial states})
represents a class of states including the well-known initial
states, such as the Werner initial state ($\left\vert
c_{1}\right\vert =\left\vert c_{2}\right\vert =\left\vert
c_{3}\right\vert =\alpha$), and Bell basis state ($\left\vert
c_{1}\right\vert =\left\vert c_{2}\right\vert =\left\vert
c_{3}\right\vert =1$).

After the coincidence of Alice and Rob, Alice stays stationary
while Rob moves with an uniform acceleration $a$. To describe the
states shared by these two relatively accelerated observers in
detail, we must use Eqs.(\ref{vacuum}) and (\ref{particle state}) to
rewrite Eq.(\ref{initial states}) in terms of Minkowski modes for
Alice, Rindler modes I for Rob and Rindler modes II for Anti-Rob,
which implies that Rob and Anti-Rob are respectively confined in
region $\mathrm{I}$ and $\mathrm{II}$. The regions $\mathrm{I}$ and
$\mathrm{II}$ are causally disconnected, and the information which
is physically accessible to the observers is encoded in the
Minkowski modes $A$ and Rindler modes $\mathrm{I}$, but the
physically unaccessible information is encoded in the Minkowski
modes $A$ and Rindler modes $\mathrm{II}$. So we must trace over the
Rindler modes $\mathrm{II}$ (modes $\mathrm{I}$) when we only
consider the Physically accessible (unaccessible) information.

\subsection{MIN shared by Alice and Rob}

We first consider the MIN between modes $A$ and $\mathrm{I}$.  By
taking the trace over the states of region $\mathrm{II}$, we obtain
\begin{eqnarray} \label{state of Alice and Rob}
\nonumber\rho_{A,I}=\frac{1}{4}
\left(
  \begin{array}{cccc}
    \frac{1+c_3}{e^{-w/T}+1} & 0 & 0 &  \frac{c_1-c_2}{(e^{-w/T}+1)^{\frac{1}{2}}} \\
    0 & (1-c_3)+\frac{1+c_3}{e^{w/T}+1} &  \frac{c_1+c_2}{(e^{-w/T}+1)^{\frac{1}{2}}}   & 0 \\
    0 &  \frac{c_1+c_2}{(e^{-w/T}+1)^{\frac{1}{2}}}  & \frac{1-c_3}{e^{-w/T}+1}  & 0 \\
    \frac{c_1-c_2}{(e^{-w/T}+1)^{\frac{1}{2}}}  & 0 & 0 &  (1+c_3)+\frac{1-c_3}{e^{w/T}+1}\\
  \end{array}
\right)
,
\end{eqnarray}
where $|mn\rangle=|m\rangle_{\mathrm{A}}|n\rangle_{\mathrm{I}}$.
For convenience to calculate the MIN, we rewrite the state
$\rho_{A,\mathrm{I}}$ in terms of Bloch representation, which is
given by
\begin{eqnarray}\label{AI Bloch representation}
\rho_{A,I}=\frac{1}{4}\left( \mathbf{1}_A\otimes\mathbf{1}_I
+c'_0\mathbf{1}_A\otimes\sigma^{(I)}_3
+\sum^3_{i=1}c'_i\sigma^{(A)}_i\otimes\sigma^{(I)}_i \right),
\end{eqnarray}
where $c'_0=\frac{-1}{(e^{w/T}+1)}$, $c'_1=\frac{c_1}{(e^{-w/T}
+1)^{\frac{1}{2}}}$, $c'_2=\frac{c_2}{(e^{-w/T}+1)^{\frac{1}{2}}}$
and $c'_3=\frac{c_3}{(e^{-w/T}+1)}$. From Eq.(\ref{MIN}), we find
that the MIN for the state $\rho_{A,I}$ is
\begin{eqnarray}\label{MIN for AI}
N(\rho_{A,I})=\frac{1}{4}\Big\{\frac{(c_1)^2}{(e^{-w/T}+1)}
+\frac{(c_2)^2}{(e^{-w/T}+1)}+\frac{(c_3)^2}{(e^{-w/T}+1)^2}
\nonumber \\
-\min[\frac{(c_1)^2}{(e^{-w/T}+1)},\frac{(c_2)^2}{(e^{-w/T}+1)},
\frac{(c_3)^2}{(e^{-w/T}+1)^2}]\Big\}.
\end{eqnarray}
Obviously, $\min\big[\frac{(c_1)^2}{(e^{-w/T}+1)},\frac{(c_2)^2}{
(e^{-w/T}+1)}, \frac{(c_3)^2}{(e^{-w/T}+1)^2}\big]$ depends on both
the coefficients $c_i$ of the states in Eq.(\ref{initial states})
and the Unruh temperature.

(i) If $|c_1|,|c_2|\geq|c_3|$ in Eq.(\ref{initial states}), the
minimum term in Eq.(\ref{MIN for AI}) is
$\frac{(c_3)^2}{(e^{-w/T}+1)^2}$. In this case, the MIN, provided
taking fixed $c_i$, decreases monotonously as the Unruh temperature
increases.

(ii) For the case of $|c_3|>\min\{|c_1|,|c_2|\}$ and both $c_1$  and
$c_2$ don't equals to 0 at the same time, if
$\min\{|c_1|,|c_2|\}\geq\frac{\sqrt{2}}{2}|c_3|$, the MIN has a
peculiar dynamics with a sudden change as the Unruh temperature
increases, i.e., $N(\rho_{A,I})$ decays quickly until
\begin{eqnarray}\label{TSC}
 T_{sc}=\frac{-w}{\ln(\frac{|c_3|^2}{min\{|c_1|^2,|c_2|^2\}}-1)},
\end{eqnarray}
and then $N(\rho_{A,I})$ decays relatively slowly. Otherwise,  the
MIN decays monotonously as the temperature increases.

(iii) Finally, if $|c_1|=|c_2|=0$, we have a monotonic decay  of
$N(\rho_{A,I})$  as the temperature increases.

The decrease of MIN means that the difference between the pre-
and post-measurement states becomes smaller, i.e., the disturbance
induced by local measurement weaken. If we understand the MIN as
some kind of correlations, this decrease means that the quantum
correlation shared by two relatively accelerated observers decreases,
i.e., less quantum resource can be used for the quantum information task
by these two observers. So the Unruh effect affects quantum communication
process by inducing the decrease of quantum resource.

\begin{figure}[htp!]
\centering
\includegraphics[width=0.5\textwidth]{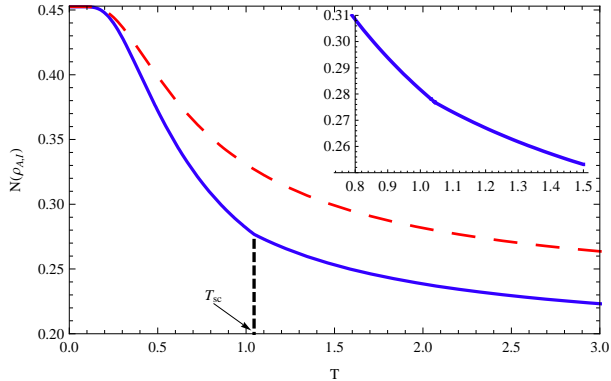}
\caption{(Color online) The MIN of state $\rho_{A,I}$  as a function
of Unruh temperature $T$. We take parameters $c_1=1$, $c_2=0.9$ and
$|c_3|\leq|c_1|,|c_2|$ for red dashed line; $c_1=0.9$, $c_2=0.85$
and $c_3=1$ for blue solid line. The insert shows the detail of
sudden change.}\label{MIN1}
\end{figure}

By taking $w=1$, the dynamical  behavior of $N(\rho_{A,I})$ is shown
in Fig. \ref{MIN1}. We find from the figure that the MIN, as the
Unruh temperature approaches to the infinite, has a limit
\begin{eqnarray}\label{limit of N}
\lim_{T\rightarrow\infty}N(\rho_{A,I})= \frac{1}{16}\{2(c_1)^2
+2(c_2)^2+(c_3)^2-\min[2(c_1)^2,2(c_2)^2,(c_3)^2]\}.
\end{eqnarray}
That is to say, as long as the initial MIN does  not equal to zero,
it can persist for arbitrary Unruh temperature.

\begin{figure}[htp!]
\centering
\includegraphics[width=0.5\textwidth]{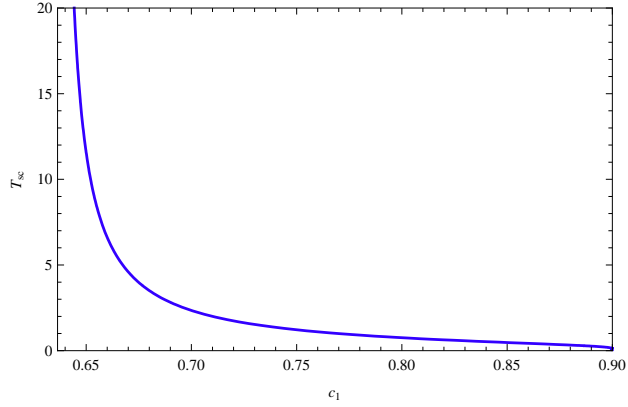}
\caption{(Color online) The $T_{sc}$ as a function of  $c_1$, here
we take $|c_1|\leq|c_2|$ and $c_3=0.9$.}\label{Tsc}
\end{figure}

We study $T_{sc}$ of Eq.(\ref{TSC}), if $|c_1|\leq|c_2|$,
by taking fixed $c_3$, we plot how the parameter $c_1$ affects it
in Fig. \ref{Tsc}, which shows that it decreases
monotonously as $c_1$ increases. That is to say, the bigger $c_1$
is, the sudden change behavior occurs earlier. And when $|c_2|\leq|c_1|$, it is
interesting to note that with the increase of $c_2$, it
decreases monotonously too.

\begin{figure}[htp!]
\centering
\includegraphics[width=0.54\textwidth]{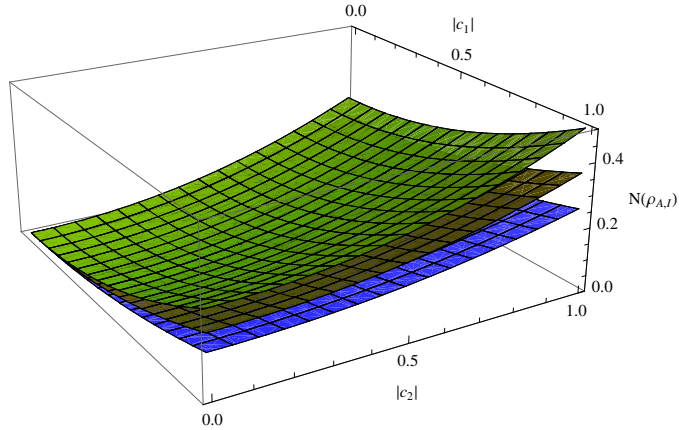}
\caption{(Color online) The MIN of state $\rho_{A,I}$ as function of
$|c_1|$ and $|c_2|$ with $|c_1|,|c_2|\geq|c_3|$. Here we take fixed
$T=0.1,1,20$ from top to bottom, respectively.}\label{MINc}
\end{figure}

In Fig. \ref{MINc}, we study how  the prepared states affect the
MIN for case (i). It is found that the $N(\rho_{A,I})$ increases
monotonously as $|c_i|$ (i=1,2) increases. And for the case that
$|c_3|>\min\{|c_1|,|c_2|\}\geq\frac{\sqrt{2}}{2}|c_3|$,  the MIN depends only on $|c_3|$ and
$\max\{|c_1|,|c_2|\}$ before $T_{sc}$, while after $T_{sc}$ it is
independent of $|c_3|$ but dependent of $|c_1|$, $|c_2|$. However,
no matter which case, the MIN increases with the increase of $|c_i|$
it depends on.

Because the MIN is introduced to describe non-locality,  its
definition is very similar to that of geometric discord. For further
understanding it, we will compare it with the maximal possible value
$\langle B_{max}\rangle$ of the Bell-CHSH expectation value and
geometric discord.

As shown in Ref.\cite{Nicolai}, the $\langle B_{max}\rangle$  for a
given state $\rho$ is determined by
\begin{eqnarray}\label{Bell expectation value}
\langle B_{max}\rangle_\rho=2\sqrt{\mu_1+\mu_2},
\end{eqnarray}
where $\mu_1$, $\mu_2$ are the two largest eigenvalues  of
$U(\rho)=TT^t$, the matrix $T=(t_{ij})$ with $t_{ij}=Tr[\rho\sigma_i\otimes\sigma_j]$.
And the geometric discord is defined as
\cite{Bor,Adesso}
\begin{eqnarray}\label{geometric discord}
D(\rho)=\frac{1}{4}(||\overrightarrow{x}||^2+||T||^2-k_{max}),
\end{eqnarray}
where $k_{max}$ is the largest eigenvalue of matrix
$K=\overrightarrow{x}\overrightarrow{x}^t+TT^t$, where
$\overrightarrow{x}=(x_i)^t$ with $x_i=Tr[\rho\sigma_i\otimes\mathbf{1}]$
and $T$ have the same definitions with Eq.(\ref{Bell expectation value}).

Using Eqs.(\ref{AI Bloch representation}), (\ref{Bell expectation
value}) and (\ref{geometric discord}), $\langle
B_{max}\rangle_{\rho_{A,I}}$ and $D(\rho_{A,I})$ are given by
\begin{eqnarray}\label{Bmax}
\langle B_{max}\rangle_{\rho_{A,I}}=2\{\frac{(c_1)^2}{(e^{-w/T}+1)}
+\frac{(c_2)^2}{(e^{-w/T}+1)}+\frac{(c_3)^2}{(e^{-w/T}+1)^2}
\nonumber \\
-\min[\frac{(c_1)^2}{(e^{-w/T}+1)},\frac{(c_2)^2}{(e^{-w/T}+1)},
\frac{(c_3)^2}{(e^{-w/T}+1)^2}]\}^{1/2},
\end{eqnarray}
and
\begin{eqnarray}\label{D}
D(\rho_{A,I})=\frac{1}{4}\{\frac{(c_1)^2}{(e^{-w/T}+1)}
+\frac{(c_2)^2}{(e^{-w/T}+1)}+\frac{(c_3)^2}{(e^{-w/T}+1)^2}
\nonumber \\
-\max[\frac{(c_1)^2}{(e^{-w/T}+1)},\frac{(c_2)^2}{(e^{-w/T}+1)},
\frac{(c_3)^2}{(e^{-w/T}+1)^2}]\},
\end{eqnarray}
respectively.

\begin{figure}[htp!]
\centering
\includegraphics[width=0.5\textwidth]{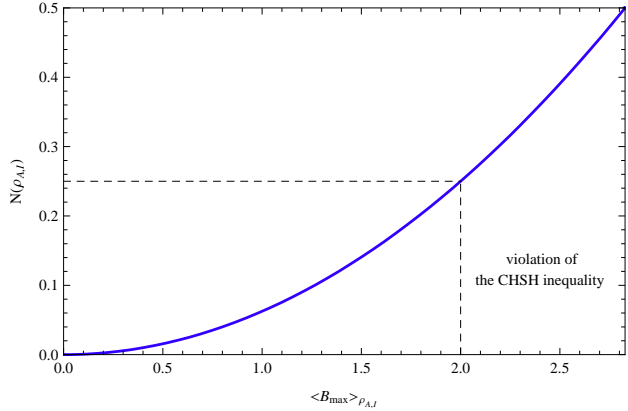}
\caption{(Color online) The MIN of state $\rho_{A,I}$  as function
of the maximally possible value of the Bell-CHSH expectation
value.}\label{NB relation}
\end{figure}

It is interesting to note that
\begin{eqnarray}
N(\rho_{A,I})=\frac{1}{16}\langle B_{max}\rangle_{\rho_{A,I}}^2.
\end{eqnarray}
We plot $N(\rho_{A,I})$ versus $\langle B_{max}\rangle_{\rho_{A,I}}$
in Fig.\ref{NB relation}, which shows that $N(\rho_{A,I})$ increases
monotonously as $\langle B_{max}\rangle_{\rho_{A,I}}$ increases and
it vanishes at zero point of $\langle B_{max}\rangle_{\rho_{A,I}}$.
It is well known that Bell inequality must be obeyed by local
realism theory, but may be violated by quantum mechanics. If we get
$\langle B_{max}\rangle_{\rho_{A,I}}>2$, it means that the violation
of Bell-CHSH inequality, which tells us that there exists nonlocal
quantum correlation. But when $\langle
B_{max}\rangle_{\rho_{A,I}}\leq2$, it doesn't mean that no quantum
correlation exists, at leat for some mixed states, which have
quantum correlation but obey the Bell inequality. So we can't be
sure that whether quantum correlation exists or not when $\langle
B_{max}\rangle_{\rho_{A,I}}\leq2$. However, the MIN, which is an
indicator of the global effect caused by locally invariant
measurement, is introduced to quantify nonlocality, and nonzero MIN
means existence of nonlocality. And form Fig.\ref{NB relation} we
see that the MIN persists for all $\langle
B_{max}\rangle_{\rho_{A,I}}$ except for zero. Thus, the MIN,
understood as some kind of correlations, is more general than the
quantum nonlocality related to violation of the Bell's inequalities.

From Eqs.(\ref{MIN for AI}) and (\ref{D}),  we can see that
the MIN is proportional to the two largest eigenvalues of the matrix $TT^t$,
while the geometric discord is proportional to the two smallest eigenvalues
of it, so we know that the MIN should be always equal or
larger than the geometric discord. In Fig. \ref{ND relation} we plot
the MIN versus the geometric discord for the Werner
($|c_1|=|c_2|=|c_3|=c$) states. It is shown that the MIN increases
monotonously as the geometric discord increases, and it is always
equal or larger than the geometric discord. So as the quantum resource,
the MIN is more robust than geometric discord.
\begin{figure}[htp!]
\centering
\includegraphics[width=0.55\textwidth]{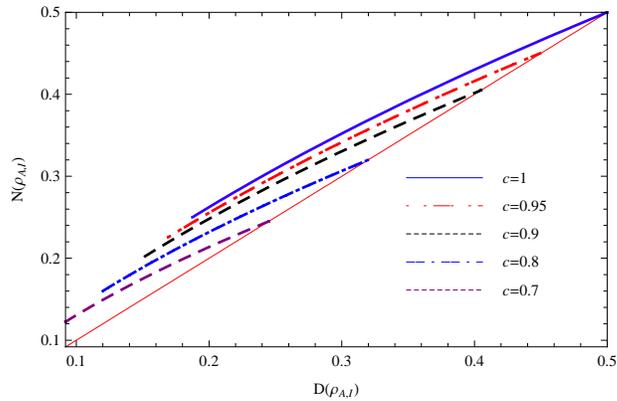}
\caption{(Color online) The MIN of state $\rho_{A,I}$ as a function
of geometric discord $D(\rho_{A,I})$ for the Werner sates, and the
red solid line represents $N(\rho_{A,I})=D(\rho_{A,I})$.} \label{ND
relation}
\end{figure}

\subsection{MIN shared by Alice and Anti-Rob}

Now we consider the MIN between modes $\mathrm{A}$  and
$\mathrm{II}$. By tracing over all modes in region $\mathrm{I}$, we
get
\begin{eqnarray}\label{AII Bloch representation}
\rho_{A,II}=\frac{1}{4}\left( \mathbf{1}_A\otimes\mathbf{1}_{II}
+c'_0\mathbf{1}_A\otimes\sigma^{(II)}_3
+\sum^3_{i=1}c'_i\sigma^{(A)}_i\otimes\sigma^{(II)}_i \right),
\end{eqnarray}
where $c'_0=\frac{1}{(e^{-w/T}+1)}$, $c'_1=\frac{c_1}{(e^{w/T}
+1)^{\frac{1}{2}}}$, $c'_2=\frac{-c_2}{(e^{w/T}+1)^{\frac{1}{2}}}$
and $c'_3=\frac{-c_3}{(e^{w/T}+1)}$. Similarly, the MIN of state
$\rho_{A,II}$ can be obtained according to Eq.(\ref{MIN}), which is
\begin{eqnarray}\label{MIN for AII}
N(\rho_{A,II})=\frac{1}{4}\{\frac{(c_1)^2}{(e^{w/T}+1)}
+\frac{(c_2)^2}{(e^{w/T}+1)}+\frac{(c_3)^2}{(e^{w/T}+1)^2}
\nonumber \\
-\min[\frac{(c_1)^2}{(e^{w/T}+1)},\frac{(c_2)^2}{(e^{w/T}+1)},
\frac{(c_3)^2}{(e^{w/T}+1)^2}]\}.
\end{eqnarray}

(i) If $|c_1|,|c_2|\geq|c_3|$, the MIN increases monotonously as the
Unruh temperature increases provided taking fixed $c_i$.

(ii) For the case of $|c_3|>\min\{|c_1|,|c_2|\}$ and both $c_1$ and
$c_2$ don't equal to 0 at the same time, if
$\min\{|c_1|,|c_2|\}\leq\frac{\sqrt{2}}{2}|c_3|$, the MIN has a
peculiar dynamics with a sudden change at $T_{sc}$
\begin{eqnarray}\label{TSC2}
T_{sc}=\frac{w}{\ln(\frac{|c_3|^2}{min\{|c_1|^2,|c_2|^2\}}-1)}.
\end{eqnarray}
Otherwise, the MIN increases monotonously with the increase of the
Unruh temperature.

(iii) Finally, if $|c_1|=|c_2|=0$, we have a monotonic increase  of
$N(\rho_{A,II})$ as the Unruh temperature increases.

\begin{figure}[htp!]
\centering
\includegraphics[width=0.5\textwidth]{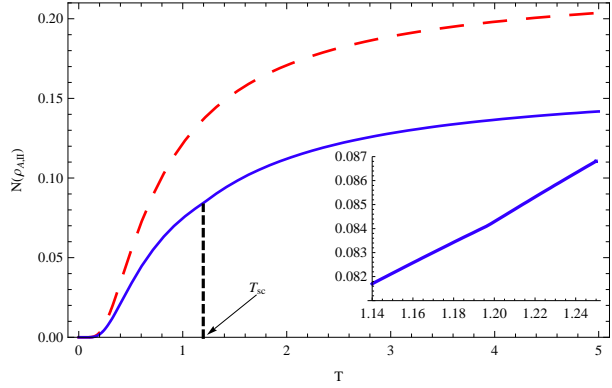}
\caption{(Color online) The MIN of state $\rho_{A,II}$  as a
function of Unruh temperature $T$.  We take parameters $c_1=1$,
$c_2=0.9$ and $|c_3|\leq|c_1|,|c_2|$ for red dashed line;
$c_1=0.9$,$c_2=0.55$ and $c_3=1$ for blue solid line. The insert
shows the detail of sudden change.}\label{MIN2}
\end{figure}

We plot $N(\rho_{A,II})$ versus the Unruh temperature in Fig.
\ref{MIN2}. It is found that the MIN, as the Unruh temperature
approaches to the infinite, is close to
\begin{eqnarray}
\lim_{T\rightarrow\infty}N(\rho_{A,II})=\frac{1}{16}\{2(c_1)^2
+2(c_2)^2+(c_3)^2-\min[2(c_1)^2,2(c_2)^2,(c_3)^2]\},
\end{eqnarray}
which is the same as $\lim_{T\rightarrow\infty}N(\rho_{A,I})$.  In
addition, as $T=0$ the MIN vanishes, which means that the
correlation between $\mathrm{A}$ and $\mathrm{II}$ is local when the
observers are inertial.

\begin{figure}[htp!]
\centering
\includegraphics[width=0.5\textwidth]{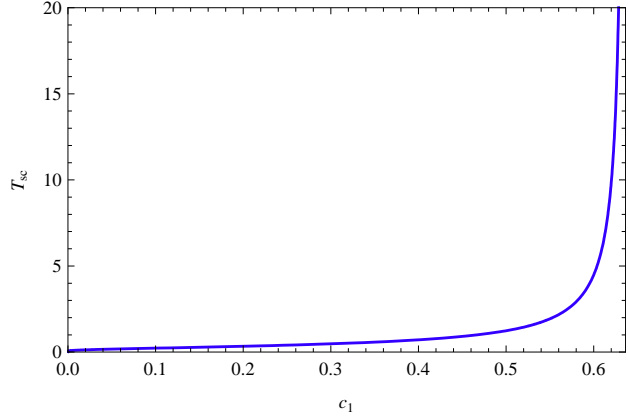}
\caption{(Color online) The $T_{sc}$ as a function  of $c_1$, here
we take $|c_1|\leq|c_2|$ and $c_3=0.9$.}\label{Tsc1}
\end{figure}

When $|c_1|<|c_2|$, by taking fixed $c_3$, we plot $T_{sc}$ as a
function of $c_1$ in Fig. \ref{Tsc1}. We learn from the figure that,
unlike the Fig.\ref{Tsc},
 $T_{sc}$ increases monotonously with the increase of $c_1$. That is to say,
the bigger $c_1$ is, the sudden change behavior occurs latter.
And when $|c_2|<|c_1|$, it is also important to note that as $|c_2|$ increases
$T_{sc}$ increases monotonously too.

\begin{figure}[htp!]
\centering
\includegraphics[width=0.55\textwidth]{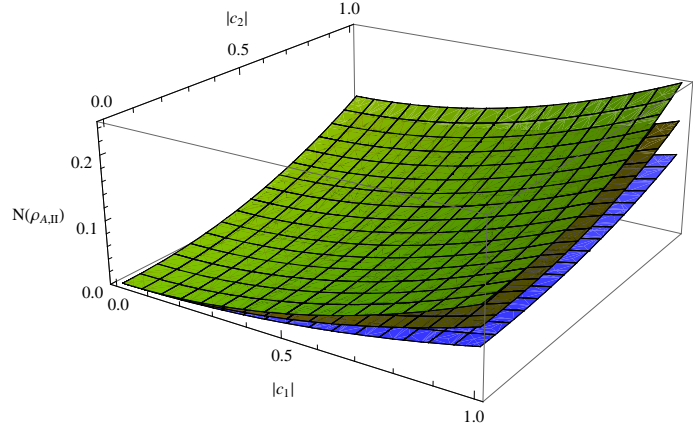}
\caption{(color online) The MIN of state $\rho_{A,II}$ as  function
of $c_1$ and $c_2$ with $|c_1|,|c_2|\geq|c_3|$. Here we take fixed
$T=\infty,2,1$ from top to bottom, respectively.}\label{MINc1}
\end{figure}

How the prepared states affect the MIN for case (i) is shown in Fig.
\ref{MINc1}, which tells us that $N(\rho_{A,II})$ increases
monotonously as $|c_i|$ ($i=1,2$) increases. And for the case that
$|c_3|>\min\{|c_1|,|c_2|\}$, the MIN is independent of $|c_3|$ but
dependent of $|c_1|$, $|c_2|$ before $T_{sc}$, while after $T_{sc}$
it depends on $|c_3|$ and $\max\{|c_1|,|c_2|\}$. However, no matter
which case, the MIN increases with the increase of $|c_i|$.

\begin{figure}[htp!]
\centering
\includegraphics[width=0.5\textwidth]{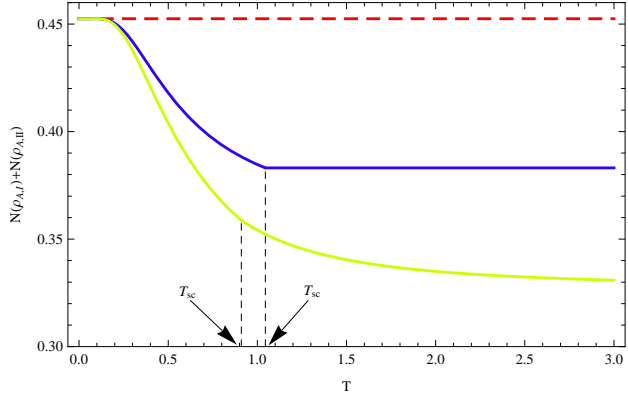}
\caption{(Color online) The sum of $N(\rho_{A,I})$  and
$N(\rho_{A,II})$ as a function of the Unruh temperature.  We take
$c_1=1,c_2=0.9$ and $|c_3|\leq|c_1|,|c_2|$ for the red dashed line;
 $c_3=1,c_1=0.9$ and $c_2=0.85$ for the blue solid line, and
 $c_3=1,c_1=0.9$ and $c_2=0.5$ for the
yellow solid line.}\label{SUMN}
\end{figure}

From the above discussions, we know that  the Unruh effect can
induce the degradation for $N(\rho_{A,I})$, but the increase for
$N(\rho_{A,II})$. However, $N(\rho_{A,I})+N(\rho_{A,II})$ has
different dynamics for different classes of states: (i) When
$|c_1|,|c_2|\geq|c_3|$, $N(\rho_{A,I})+N(\rho_{A,II})$ is
independent of the Unruh temperature. That is to say,
$N(\rho_{A,I})+N(\rho_{A,II})$ is a constant versus the Unruh
temperature for this class of states; (ii) When
$|c_3|>\min\{|c_1|,|c_2|\}\geq\frac{\sqrt{2}}{2}|c_3|$, with the
increase of the Unruh temperature $N(\rho_{A,I})+N(\rho_{A,II})$
decreases monotonously until
\begin{eqnarray}
T_{sc}=\frac{-w}{\ln(\frac{|c_3|^2}{min\{|c_1|^2,|c_2|^2\}}-1)},
\end{eqnarray}
and from then on it remains constant;  And (iii) when
$\min\{|c_1|,|c_2|\}<\frac{\sqrt{2}}{2}|c_3|$,
$N(\rho_{A,I})+N(\rho_{A,II})$ decays quickly until
\begin{eqnarray}
T_{sc}=\frac{w}{\ln(\frac{|c_3|^2}{min\{|c_1|^2,|c_2|^2\}}-1)},
\end{eqnarray}
and after that it decays relatively slowly.  We plot these dynamical
behaviors in Fig. \ref{SUMN}.

\section{conclusions}

The effect of the prepared states and Unruh temperature  on the MIN
of Dirac fields was investigated and the following new properties
were found: (i) The MIN $N(\rho_{A,I})$ for the X-type states
decreases as the Unruh temperature increases, but $N(\rho_{A,II})$
increases with the increase of the Unruh temperature. (ii) For fixed
Unruh temperature, it is found that the MIN always increases as
$|c_i|$ $(i=1,2,3)$ increases, and it takes the maximal value for
the Bell basis states. (iii) Both $N(\rho_{A,I})$ and
$N(\rho_{A,II})$ have a peculiar dynamics with a sudden change at
$T_{sc}$ provided $c_i$ appropriately chosen, the $T_{sc}$ for
$N(\rho_{A,I})$ decreases as $c_i$ increases, while it is contrary
for $N(\rho_{A,II})$. (iv) The MIN is more general than the quantum
nonlocality related to violation of Bell's inequalities. Besides, it
increases as the geometric discord increases, and it is always equal
or larger than the geometric discord. And (v)
$N(\rho_{A,I})+N(\rho_{A,II})$ has three kinds of dynamics: (a) When
$|c_1|,|c_2|\geq|c_3|$, it is independent of the Unruh temperature;
(b) When $|c_3|>\min\{|c_1|,|c_2|\}\geq\frac{\sqrt{2}}{2}|c_3|$,
with the increase of the Unruh temperature it decreases monotonously
until
$T_{sc}=\frac{-w}{\ln(\frac{|c_3|^2}{min\{|c_1|^2,|c_2|^2\}}-1)}$,
and from then on it remains constant;  And (c) when
$\min\{|c_1|,|c_2|\}<\frac{\sqrt{2}}{2}|c_3|$, with the increase of
the Unruh temperature, it decays quickly until
$T_{sc}=\frac{w}{\ln(\frac{|c_3|^2}{min\{|c_1|^2,|c_2|^2\}}-1)}$,
and after that it decays relatively slowly.

Here we just simply discuss the relation between the MIN and the
maximal expectation values of CHSH inequality and geometric discord.
More detailed study of this relation can help us to ont only
understand the MIN more clearly, but also distinguish difference of
quantum resource based on different correlation measurements. Such
topics are left for future research.

\begin{acknowledgments}
This work was supported by the  National Natural Science Foundation
of China under Grant No. 11175065, 10935013; the National Basic
Research of China under Grant No. 2010CB833004; the SRFDP under
Grant No. 20114306110003; PCSIRT, No. IRT0964; the Hunan Provincial
Natural Science Foundation of China under Grant No 11JJ7001;  and
Construct Program of the National Key Discipline.
\end{acknowledgments}

\end{document}